\documentclass[12pt,a4paper]{article}
\usepackage{graphicx}
\usepackage{times}
\title{\bf The WR 140 periastron passage 2009: first results from MONS and other optical sources}
\author{R. Fahed$^1$, A. F. J. Moffat$^1$, J. Zorec$^2$, T. Eversberg$^3$, A. N. Chen\'e$^4$,\\ F. Alves*, W. Arnold*, T. Bergmann*, L. F. Gouveia Carreira*, F. Marques Dias*,\\ A. Fernando*, J. Sanchez Gallego*, T. Hunger*, J. H. Knapen*, R. Leadbeater*, \\ T. Morel*, G. Rauw*, N. Reinecke*, J. Ribeiro*, N. Romeo*, \\ E. M. dos Santos*, L. Schanne*, O. Stahl*, Ba. Stober*, Be. Stober*, \\ N. G. Correia Viegas*, K. Vollmann*, M. F. Corcoran*, S. M. Dougherty*,\\ J. M. Pittard*, A. M. T. Pollock*, and P. M. Williams* \\
\vspace{1cm}\\
\normalsize $^1$ Universit\'e de Montr\'eal, Montr\'eal, Canada\\ 
\normalsize $^2$ Institut d’astrophysique de Paris, Paris, France \\
\normalsize $^3$ Schn\"orringen Telescope Science Institute, K\"oln, Germany \\
\normalsize $^4$ Herzberg Institute of Astrophysics, Victoria, Canada \\
\normalsize * MONS pro-am collaboration 
\\
\\
\normalsize Published in proceedings of \\
\normalsize"Stellar Winds in Interaction", editors T. Eversberg and J.H. Knapen. \\ 
\normalsize Full proceedings volume is available on http://www.stsci.de/pdf/arrabida.pdf
}
\date{\mbox{}}
\begin{document}
\maketitle
%
%
\def\bull{\vrule height .9ex width .8ex depth -.1ex}
\makeatletter
\def\ps@plain{\let\@mkboth\gobbletwo
\def\@oddhead{}\def\@oddfoot{\hfil\tiny\bull\quad
Workshop ``Stellar Winds in Interaction'' Convento da Arr\'abida, 2010 May 29 - June 2\quad\bull}%
\def\@evenhead{}\let\@evenfoot\@oddfoot}
\makeatother
%
%
\def\beginrefer{\section*{References}%
\begin{quotation}\mbox{}\par}
\def\refer#1\par{{\setlength{\parindent}{-\leftmargin}\indent#1\par}}
\def\endrefer{\end{quotation}}
%
%
{\noindent\small{\bf Abstract:} 
We present the results from the spectroscopic follow-up of WR140 (WC7 + O4-5) during its last periastron passage in January 2009. This object is known as the archetype of colliding wind binaries and has a relatively large period ($\sim8$\,years) and eccentricity ($\sim0.89$). We provide updated values for the orbital parameters, new estimates for the WR and O star masses and new constraints on the mass-loss rates.
}
%
%
\section{Introduction}
WR140 is a very eccentric WC7+O5 colliding-wind binary (CWB) system with an eccentricity of 0.89 and a long period of 7.94 years. It is also the brightest Wolf-Rayet star in the northern hemisphere and is considered as the archetype of CWB. We present here the results from a spectroscopic follow-up, unique in time coverage and resolution. The observation campaign was a worldwide collaboration involving professional and amateur astronomers and took place during a period of 4 months around periastron passage in January 2009.

\section{Observations}
Among the amateur data, we first have the MONS project: under the leadership of Thomas Eversberg, a German amateur astronomer who has founded his own astronomical observatory (the Schn\"orringen Telescope Science Institute), a LHIRES spectrograph was installed on a telescope now owned by the Instituto de Astrof\'\i sica de Canarias (IAC), but previously by the University of Mons (Belgium) and which is used mainly for pedagogical purposes. The 50\,cm telescope is located at the Teide observatory of the IAC in Tenerife. During four months, data have been acquired with this instrument. Other amateurs contributed using their own personal instruments in Portugal, Germany and England.

The professional data were obtained with the echelle spectrograph SOPHIE at the Observatoire de Haute Provence (OHP), at the Dominion Astrophysical Observatory (DAO) and at the Observatoire du Mont M\'egantic (OMM). A list of the data sources and characteristics of the campaign is presented Table~\ref{sources}.

\begin{table}
\caption{List of the different sources in the 2009 campaign. \label{sources}}
\small
\begin{center} 
\begin{tabular}{c|c|c|c|c}
Observatory & Dates & Wavelength Range (\AA) & Resolution (\AA / pixel) & Number of spectra  \\
\hline
Tenerife & 1.12.08 - 23.03.09 & 5530-6000 & 0.35 & 34  \\
\hline
OHP & 12.12.08 - 23.3.09 & 4000-6800 & 0.01 & 63  \\
\hline
DAO & 22.4.08 - 9.1.09 & 5350-5900 & 0.37 & 13  \\
\hline
OMM & 5.7.09 - 8.8.09 & 4500-6000 & 0.63 & 18  \\
\hline
Robin Leadbeater & 10.12.07 - 20.3.09 & 5600-6000 & 0.68 & 38  \\
\hline
Berthold Stober & 26.8.08 - 29.2.09 & 5500-6100 & 0.53 & 12  \\
\end{tabular}
\end{center}
\end{table}

\section{Radial velocities}
The WR star radial velocities were measured by cross correlation with a reference spectrum and the O star radial velocities by measuring the centroid of the photospheric absorption lines (see Fig.~1). We notably find a higher eccentricity than previously published ($e=0.896 \pm 0.002$ cf. 0.881 $\pm$ 0.005 from  Marchenko et al. 2003 = M03) and update the value for the period (2896.5 $\pm$ 0.7\,d instead of 2899.0 $\pm$ 1.3\,d).

\begin{figure}
  \centering
\includegraphics[width=0.90\textwidth]{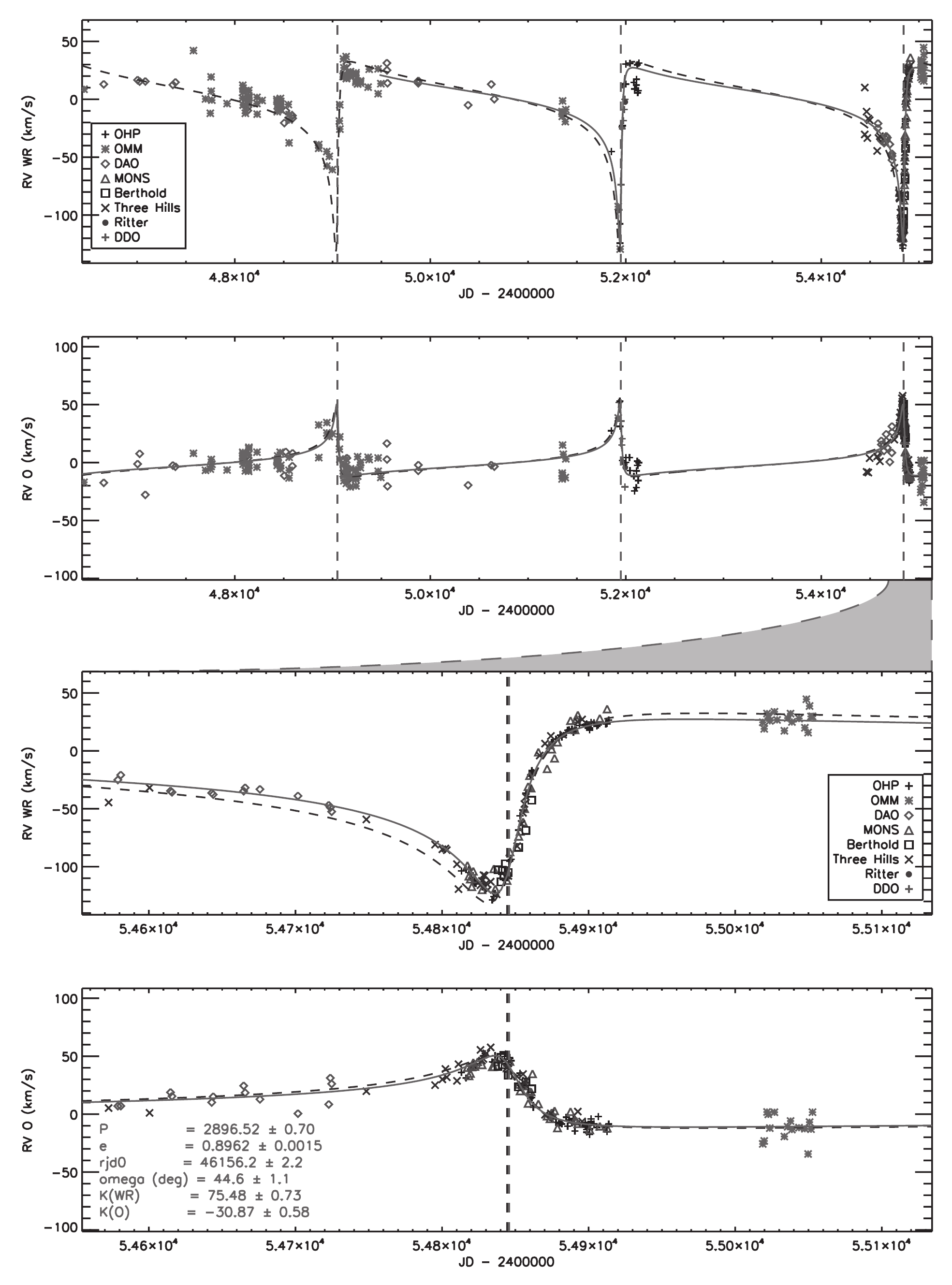}
  \caption{(Top two panels) Measured radial velocities of the WR star and of the O star together with the fit for the orbital solution (full line). We included data from the last periastron campaign in 2001 (M03). The black dashed line is the orbital solution from M03. The dashed vertical lines show the position of the periastron passage. (Bottom two panels) Same plots but zoomed in on the 2009 campaign. The best fit parameters are indicated in grey.}
  \label{RV}
\end{figure}

\section{Excess emission}

\begin{figure}
  \centering
\includegraphics[width=0.90\textwidth]{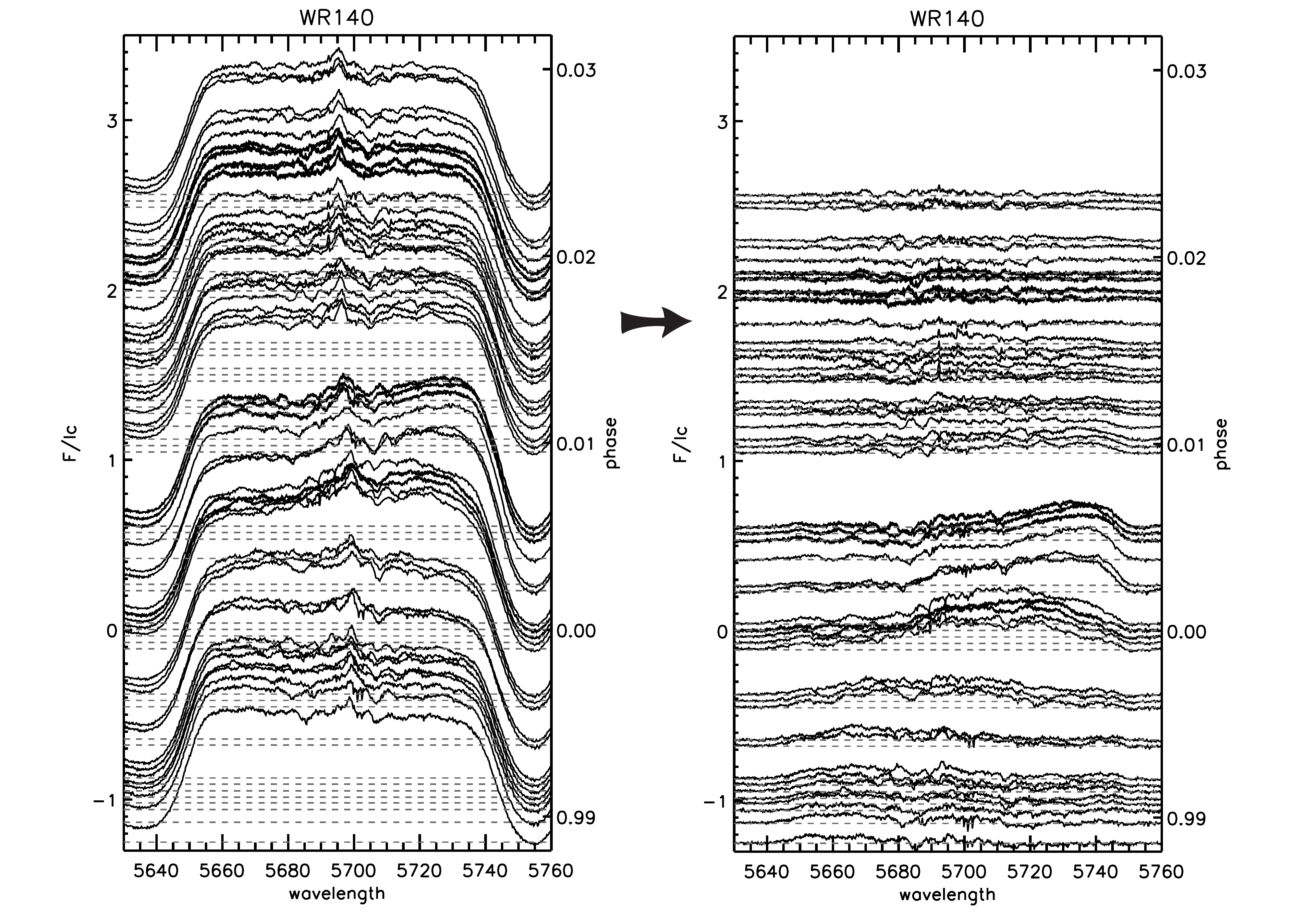}
  \caption{(Left) The C{\sc iii} 5696 flat top line as a function of the orbital phase. (Right) Excess emission as a function of the phase, obtained by substraction of a reference profile, unaffected by wind collision.}
  \label{XS}
\end{figure}

\begin{figure}
  \centering
\includegraphics[width=0.70\textwidth]{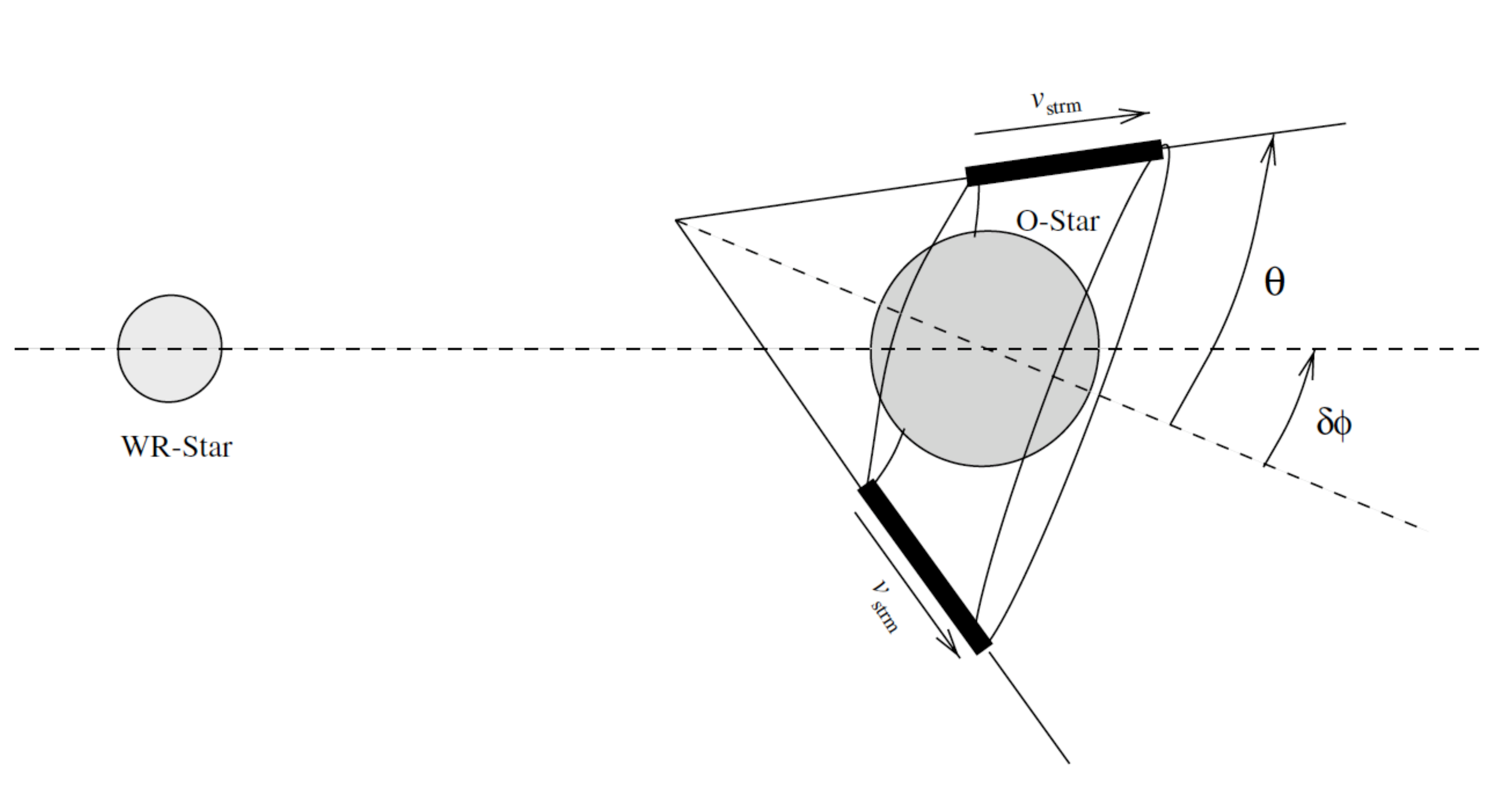}
  \caption{Schematic view of the geometric model by Luehrs (1997). The full
width and radial velocity of the excess will then be given by : FW$_{\rm ex}=C_1 + 2\,v_{\rm strm}\sin(\theta) \sqrt{1-\sin^2(i)\cos^2(\phi - \delta \phi)}$ and RV$_{\rm ex}=C_2 - v_{\rm strm}\cos(\theta)\sin(i)\cos(\phi - \delta \phi) $}
  \label{LuhrsGeom}
\end{figure}

\begin{figure}
\begin{minipage}{10cm}
\centering
\includegraphics[width=10cm]{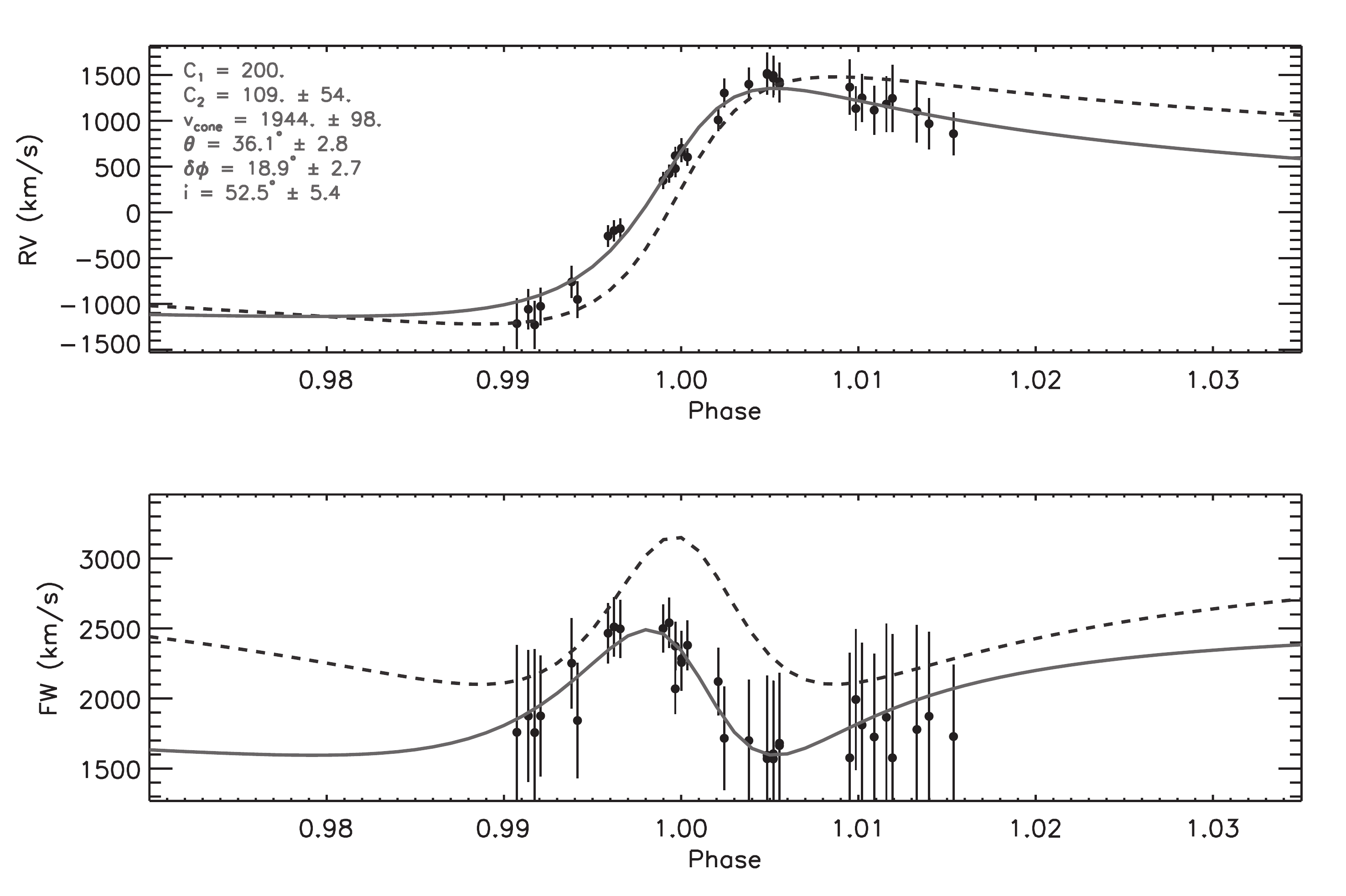}
\caption{Fit of the radial velocity and width of the excess using the Luehrs (1997) model (full grey line). The black
dashed line shows the solution from M03. \label{XSfit}}
\end{minipage}
\hfill
\begin{minipage}{6cm}
\centering
\includegraphics[width=6cm]{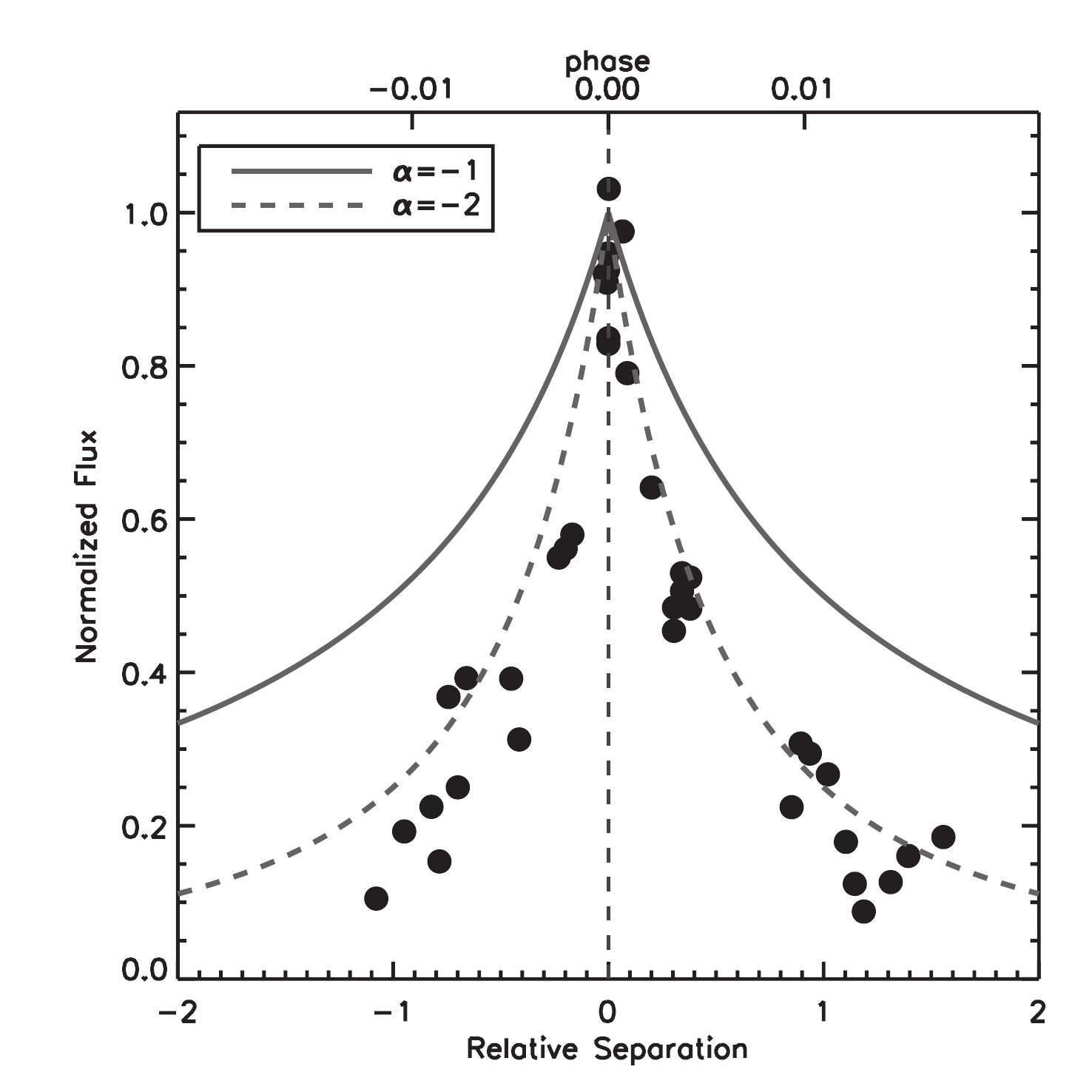}
  \caption{Normalized flux of the excess as a function of the relative separation
of the two stars ( $[d-d_{\rm min}]/d_{\rm min}$ ). The full line shows a $d^{-1}$ dependancy, expected for an adiabatic emission process. The dashed line shows a $d^{-2}$ dependency, possibly more in line with an isothermal process. \label{XSeqwidth}}
\end{minipage}
\end{figure}

The presence of a shock cone around the O star induces an excess emission that we measured on the C{\sc iii} 5696 flat-top line. This excess emission appears first, just before periastron passage, on the blue side of the line, and then moves quickly to the red side, just after periastron passage, before it disappears (Fig.~\ref{XS}). We fitted the radial velocity and the width of this excess as a function of orbital phase using a simple geometric model  (Luehrs 1997) taking into account the half opening angle of the shock cone $\theta$, the velocity of the fluid along the cone $v_{\rm strm}$, the orbital inclination $i$ and an angular shift due to Coriolis forces $\delta \phi$ (see Fig.~\ref{LuhrsGeom}). The result of this fit is shown in Fig.~\ref{XSfit}. We find a value for the inclination of 52$^{\circ}$$\pm$8$^{\circ}$ (cf. 58$^{\circ}$$\pm$5$^{\circ}$ from  Dougherty et al. 2005), which yields the following estimate for the stellar masses : $M_{\rm WR}$ = 18.4$\pm$1.8\,$M_{\odot}$ and $M_{\rm O}$ = 45.1$\pm$4.4\,$M_{\odot}$ (cf. 19\,$M_{\odot}$ and 50\,$M_{\odot}$ from M03). From the half opening angle of the shock cone (Canto et al. 1996), we also find a wind momentum ratio $\eta$ = 0.028$\pm$0.009.

\section{Conclusion}

The 2009 periastron campaign on WR140 provided updated values for the orbital parameters, new estimates for the WR and O star masses and new constraints on the mass-loss rates. However, our capability to measure the shock cone parameters with confidence and to understand its underlying physics is limited by the over simplistic approach of our model. A more sophisticated theoretical investigation should be done. Meanwhile, the $d^{-2}$ dependency of the excess, shown in Fig.~\ref{XSeqwidth}, strongly suggests that some kind of isothermal process is involved here. Links with observations in other spectral domains (X-ray, infrared, and radio) will certainly provide valuable clues about the physics. Finally, we will attempt to isolate the WR spectrum from the O-star spectrum from our data in order to identify the spectral type of the latter more precisely. We also have some photometric and spectropolarimetric data to analyse to complete our view of this system.

%
%
 
%
%
\footnotesize

\beginrefer

\refer Canto, J., Raga, A.C. \& Wilkin, F.P., 1996, ApJ, 469, 729

\refer Dougherty, S.M., Beasley, A.J., Claussen, M.J., Zauderer, B.A. \& Bolingbroke, N.J., 2005, ApJ, 623, 447–459.

\refer Luehrs, S., 1997, PASP, 109, 504–513.

\refer Marchenko, S.V., Moffat, A.F.J., Ballereau, D., Chauville, J., Zorec, J., Hill, G.M., Annuk, K., Corral, L.J., Demers, H., Eenens, P.R.J., Panov, K.P., Seggewiss, W., Thomson, J.R. \& Villar-Sbaffi, A., 2003, ApJ, 596, 1295–1304.

\endrefer

\newpage
\qquad

\end{document}